\documentclass[prl,twocolumn,showpacs,preprintnumbers,amsmath,amssymb]{revtex4}

\usepackage{natbib}
\usepackage{bibentry}
\usepackage{graphicx}
\usepackage{dcolumn}
\usepackage{bm}
\newcommand{\bra}[1]{\langle#1|}
\newcommand{\ave}[1]{\langle#1\rangle}
\newcommand{\modul}[1]{|#1|}
\newcommand{\ket}[1]{|#1\rangle}

\begin{document}

\preprint{APS/123-QED}

\title{Dynamics of Non Classically Reproducible Entanglement}

\author{Bruno Bellomo}
\author{Rosario Lo Franco}
\email{lofranco@fisica.unipa.it}
 \homepage{http://www.fisica.unipa.it/~lofranco}
\author{Giuseppe Compagno}
\affiliation{CNISM and Dipartimento di Scienze Fisiche ed Astronomiche,
Universit\`{a} di Palermo, via Archirafi 36, 90123 Palermo, Italy}

\date{\today}

\begin{abstract}
We investigate when the quantum correlations of a bipartite system, under the influence of environments with memory, are not reproducible with certainty by a classical local hidden variable model. To this purpose, we compare the dynamics of a Bell inequality with that of entanglement, as measured by concurrence. We find time regions when Bell inequality is not violated even in correspondence to high values of concurrence (up to $\approx 0.8$). We also suggest that these results may be observed by adopting a modification of a recent experimental optical setup. These findings indicate that even highly entangled systems cannot be exploited with certainty in contexts where the non classical reproducibility of quantum correlations is required.
\end{abstract}

\pacs{03.65.Ud, 03.65.Yz, 42.50.Dv}

\maketitle

In contexts such as quantum computation and quantum information, working protocols require for their implementation systems that present the peculiar quantum correlations characterized by entanglement \cite{nielsenchuang,bennett2000Nature}. However, unavoidable interaction of realistic systems with their environment gives rise to an increase of mixedness of the state of the systems and to a decrease of the degree of entanglement with time. Entanglement can even disappear completely at finite time (entanglement sudden death) \cite{yu2004PRL}. This motivates the interest in considering conditions and methods that maintain the systems entangled as long as possible. Among the conditions that effectively increase the entanglement usefulness time there are the use of non-Markovian environments, where entanglement revivals are possible \cite{bellomo2007PRL,bellomo2008PRA}, of quantum Zeno effect \cite{maniscalco2008PRL}, or of structured environments, which can give rise to entanglement trapping \cite{bellomo2008arxiv,wang2008arxiv}.

It has been however shown that there exist entangled bipartite mixed states whose correlations can be reproduced by a local hidden variable model, that is by classical systems \cite{werner1989PRA}, although they may still display ``some hidden non-locality'' \cite{masanes2008PRL}. This indicates that, for mixed states, which are in practice the ones always encountered, a given value of entanglement by itself does not imply that their correlations cannot be classically reproduced with certainty. Hereafter, we refer to quantum correlations that are certainly non reproducible by a classical local model as inherently nonlocal correlations (INLCs). Because of the necessity of INLCs for device-independent and security-proof quantum key distribution protocols \cite{acin2006PRL,gisin2007natphoton} and their relevance for quantum computation, it appears crucial to have indicators of their presence. One of such indicators is obviously given, for bipartite systems, by a Bell function and the presence of INLCs unambiguously identified by it violating a Bell inequality \cite{bell,clauser}. From another side, the question is to determine when quantum traits, linked to entanglement and suitable for quantum computation, are not classically reproducible with certainty.

The aim of this paper is thus to consider one of the cases where the time when the entanglement is present can be extended and to compare it with the time regions when INLCs are present. The system we shall consider is that of two qubits in environments with memory (non-Markovian). These systems present in general entanglement revivals and one expects that, with an appropriate choice of parameters, also revivals of INLCs occur. Finally, we shall examine if the conditions when this happens can be obtained within the current experimental technologies.

We take two independent non-causally connected qubits, each interacting with a distinct, but identical bosonic reservoir at zero temperature. Let $\{\ket{0},\ket{1}\}$ be the basis states of the qubit and the operator $\mathcal{O}$ a pseudo-spin observable with eigenvalues $\pm1$, defined as $\mathcal{O}=\textbf{O}\cdot\bm{\sigma}$, where $\textbf{O}\equiv(\sin\theta\cos\phi,\sin\theta\sin\phi,\cos\theta)$ is the unit vector indicating a direction in the pseudo-spin space and $\bm{\sigma}=(\sigma_1,\sigma_2,\sigma_3)$ the Pauli matrices vector. The expression of $\mathcal{O}$ in terms of the basis states is $\mathcal{O}(\theta,\phi)=\cos\theta(\ket{1}\bra{1}-\ket{0}\bra{0})+\sin\theta(e^{i\phi}\ket{1}\bra{0}+e^{-i\phi}\ket{0}\bra{1})$. The Clauser-Horne-Shimony-Holt (CHSH) form of the Bell function associated to the two-qubit state $\hat{\rho}$ for the operator $\mathcal{O}$ is \cite{clauser}
{\setlength\arraycolsep{1pt}\begin{eqnarray}\label{bellfunction}
B(\hat{\rho})&=&|\ave{\mathcal{O}_1(\theta_1,\phi_1)\mathcal{O}_2(\theta_2,\phi_2)}
-\ave{\mathcal{O}_1(\theta_1,\phi_1)\mathcal{O}_2(\theta'_2,\phi_2)}|\nonumber\\
&+&\ave{\mathcal{O}_1(\theta'_1,\phi_1)\mathcal{O}_2(\theta_2,\phi_2)}
+\ave{\mathcal{O}_1(\theta'_1,\phi_1)\mathcal{O}_2(\theta'_2,\phi_2)},\nonumber\\
\end{eqnarray}}where $\ave{\mathcal{O}_1\mathcal{O}_2}=\mathrm{Tr}\{\hat{\rho}\mathcal{O}_1\mathcal{O}_2\}$ is the correlation function, with the index $S=1,2$ referring to the $S$-th qubit. If, given the state $\hat{\rho}$, a set of angles $\{\phi_1,\phi_2\}$ and $\{\theta_1,\theta'_1,\theta_2,\theta'_2\}$ exists such that the CHSH-Bell inequality $B(\hat{\rho})\leq2$ is violated, the correlations cannot be simulated by any classical local model and are nonlocal. Such a set of angles always exists for pure entangled states but generally not for mixed states \cite{gisin1991PLA}.

The Bell function at time $t$ is obtained from the two-qubit state $\hat{\rho}(t)$. In our system this can be determined, for any initial state, by the knowledge of the single-qubit dynamics, whose exact solution is available when the reservoir is at zero temperature and has a memory (non-Markovian) \cite{bellomo2007PRL}. In this case, the single qubit-reservoir evolution is represented by the quantum map, known as amplitude decay channel \cite{nielsenchuang},
\begin{eqnarray}\label{amplitudedecaychannel}
&\ket{0_S}\otimes\ket{0_R}\rightarrow\ket{0_S}\otimes\ket{0_R},&\nonumber\\
&\ket{1_S}\otimes\ket{0_R}\rightarrow\sqrt{1-p}\ket{1_S}\otimes\ket{0_R}+\sqrt{p}\ket{0_S}\otimes\ket{1_R},&
\end{eqnarray}
where $\ket{0_S},\ket{1_S}$ and $\ket{0_R},\ket{1_R}$ are respectively the states of the qubit $S$ and of the reservoir $R$, and $p$ is the decay probability given by \cite{petru}
\begin{equation}\label{decayprobability}
p=1-\mathrm{e}^{-\lambda t}\left[\cos\left(\frac{dt}{2}\right)+\frac{\lambda}{d}\sin\left(\frac{dt}{2}\right)\right]^2,
\end{equation}
with $d=\sqrt{2\Gamma \lambda-\lambda^2}$. In Eq.~(\ref{decayprobability}), $\lambda$ defines the spectral width of the coupling and is connected to the reservoir correlation time $\tau_R$ by $\lambda\approx\tau_R^{-1}$, $\Gamma$ represents the qubit excited state decay rate in the Markovian limit of flat spectrum and is linked to the system (qubit) relaxation time $\tau_S$ by $\Gamma\approx\tau_S^{-1}$. We consider here the strong coupling regime defined by $\lambda/\Gamma<2$, where the reservoir correlation time is larger than the qubit relaxation time and memory effects become relevant.

The two-qubit dynamics can be analyzed in a simple way for any initial two-qubit state \cite{bellomo2007PRL}; however, here we shall restrict the analysis to the pure Bell-like initial states $\hat{\rho}_\Phi(0)=\ket{\Phi}\bra{\Phi}$ and $\hat{\rho}_\Psi(0)=\ket{\Psi}\bra{\Psi}$, where
\begin{equation}
\ket{\Phi}=\alpha\ket{01}+e^{i\delta}\beta\ket{10},\quad \ket{\Psi}=\alpha\ket{00}+ e^{i\delta}\beta\ket{11},
\end{equation}
with $\alpha,\beta$ non-negative reals and $\alpha^2+\beta^2=1$. These states belong to the class of X states whose density matrix $\hat{\rho}_X$, in the standard basis $\mathcal{B}=\{\ket{1}\equiv\ket{11},\ket{2}\equiv\ket{10},\ket{3}\equiv\ket{01},\ket{4}\equiv\ket{00}\}$, has an X structure. This structure has been shown to persist during the time evolution due to the map of Eq.~(\ref{amplitudedecaychannel}) and has the form \cite{bellomo2007PRL}
\begin{equation}\label{Xstates}
   \hat{\rho}_X (t)= \left(
\begin{array}{cccc}
  \rho_{11}(t) & 0 & 0 & \rho_{14}(t)  \\
  0 & \rho_{22}(t) & \rho_{23}(t) & 0 \\
  0 & \rho_{23}(t)^* & \rho_{33}(t) & 0 \\
  \rho_{14}(t)^* & 0 & 0 & \rho_{44}(t) \\
\end{array}
\right).
\end{equation}
In our system the initially pure states $\hat{\rho}_\Phi(0)=\ket{\Phi}\bra{\Phi}$, $\hat{\rho}_\Psi(0)=\ket{\Psi}\bra{\Psi}$ become, during the time evolution, the mixed states $\hat{\rho}_\Phi(t)$, $\hat{\rho}_\Psi(t)$ whose matrix elements as explicit functions of time $t$ and of probability amplitude $\alpha$ have been previously reported \cite{bellomo2007PRL}.

Since our goal is to find the existence of regions where $B>2$, it is strategic to maximize, by an appropriate choice of the angles, the Bell function $B(\hat{\rho})$ of Eq.~(\ref{bellfunction}). For any state $\hat{\rho}_X (t)$ the standard procedure \cite{gisin1991PLA} determines these angles as $\{\phi_1,\phi_2\}=\{(k+k')\pi-\delta_{14}-\delta_{23},(k-k')\pi-\delta_{14}+\delta_{23}\}$, where $k,k'$ are integer numbers and $\delta_{14}$, $\delta_{23}$ respectively the initial phases of $\rho_{14}(t)$, $\rho_{23}(t)$, in addition $\{\theta_1,\theta'_1,\theta_2(t),\theta'_2(t)\}=
\{0,\frac{\pi}{2},\arctan\frac{\mathcal{Q}(t)}{|\mathcal{P}(t)|},\pi-\theta_2(t)\}$. With this choice the maximum of $B$ is
\begin{equation}\label{Bmaxgeneral}
B_\mathrm{max}(\hat{\rho}_X(t))=2\sqrt{\mathcal{P}^2(t)+\mathcal{Q}^2(t)},
\end{equation}
where
\begin{eqnarray}\label{PandQ}
\mathcal{P}(t)&=&\rho_{11}(t)+\rho_{44}(t)-\rho_{22}(t)-\rho_{33}(t),\nonumber\\ \mathcal{Q}(t)&=&2(\modul{\rho_{14}(t)}+\modul{\rho_{23}(t)}).
\end{eqnarray}
The expression for $B_\mathrm{max}(\hat{\rho}_X(t))$ given by Eq.~(\ref{Bmaxgeneral}) coincides with the one which would be obtained using the formal Horodecki criterion \cite{horodecki1995PLA}.

\begin{figure}
\begin{center}
{\includegraphics[width=0.3\textwidth, height=0.19\textheight]{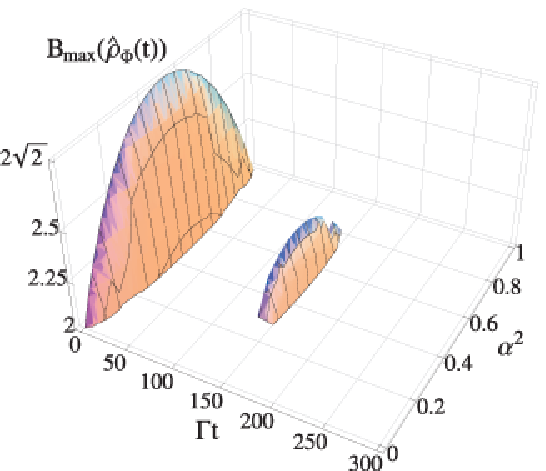}\vspace{0.5cm}
\includegraphics[width=0.3\textwidth, height=0.19\textheight]{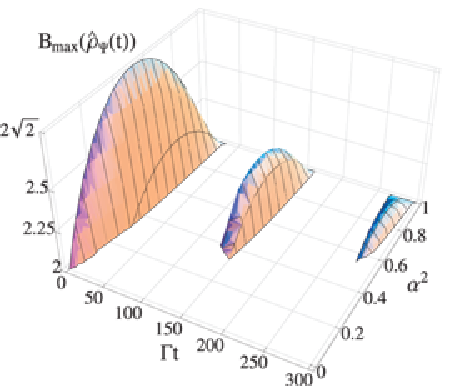}}
\end{center}
\caption{\label{fig:Bellislands}Maximum of Bell function, $B_\mathrm{max}$, in terms of the initial degree of entanglement (represented by $\alpha^2$) and of the dimensionless time $\Gamma t$ for the initial Bell-like states $\ket{\Phi}=\alpha\ket{01}+\beta\ket{10}$ (upper plot) and $\ket{\Psi}=\alpha\ket{00}+\beta\ket{11}$ (lower plot) at $\lambda/\Gamma=10^{-3}$. The parts of the graphs emerging after the first collapse under the classical threshold value 2, ``Bell islands'', correspond to a return of INLCs.}
\end{figure}
In Fig.~\ref{fig:Bellislands} the regions where $B_\mathrm{max}>2$ are reported as function of time and of the initial degree of entanglement (represented by $\alpha^2$). The figure clearly displays regions of revivals of CHSH-Bell inequality violations (Bell Islands). These regions correspond to a return, after finite intervals during which $B_\mathrm{max}\leq 2$, of INLCs at space-like distances via local qubit-environment interaction.
\begin{figure}
\begin{center}
{\includegraphics[width=0.35\textwidth, height=0.16\textheight]{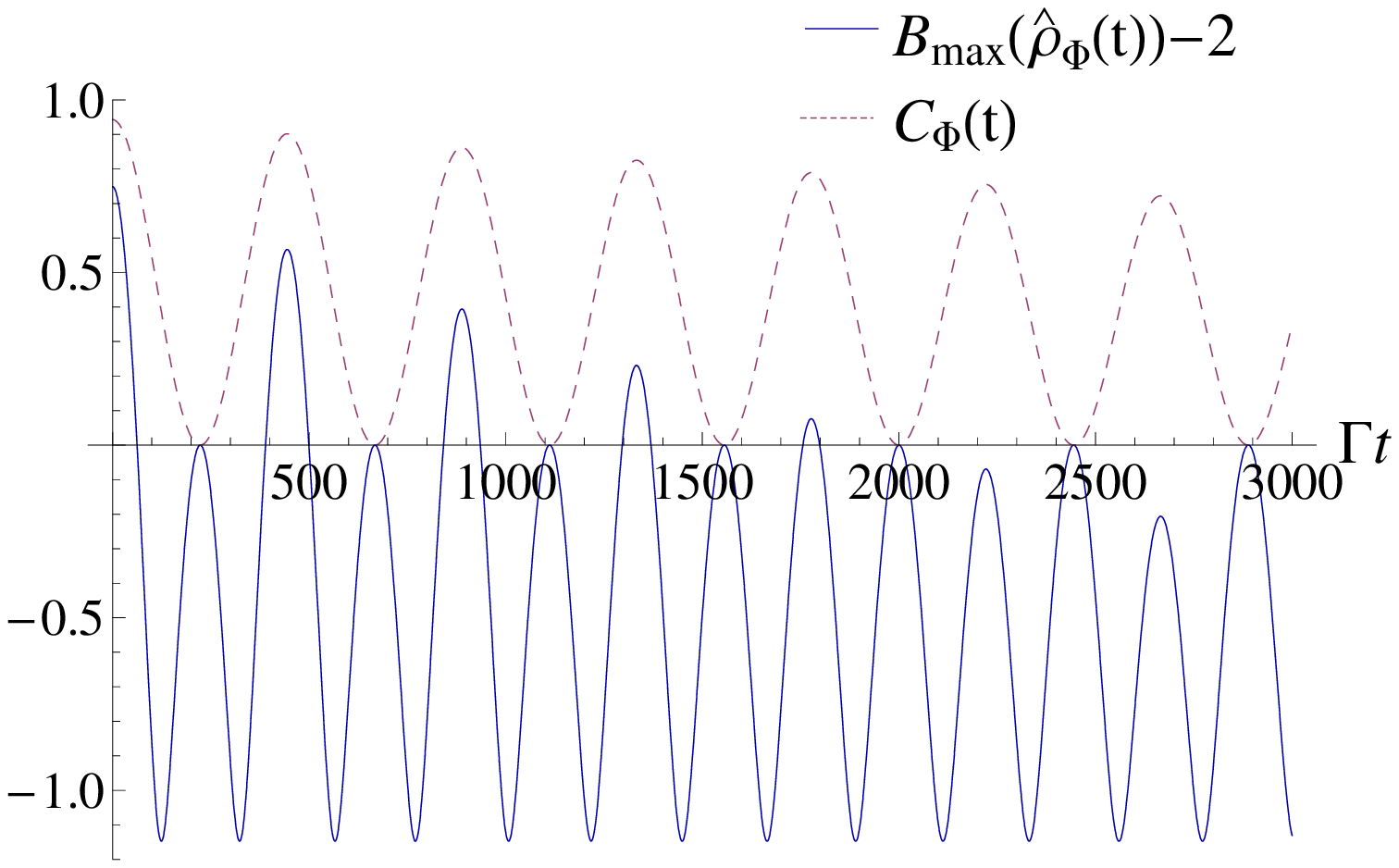}\vspace{0.3cm}
\includegraphics[width=0.35\textwidth, height=0.16\textheight]{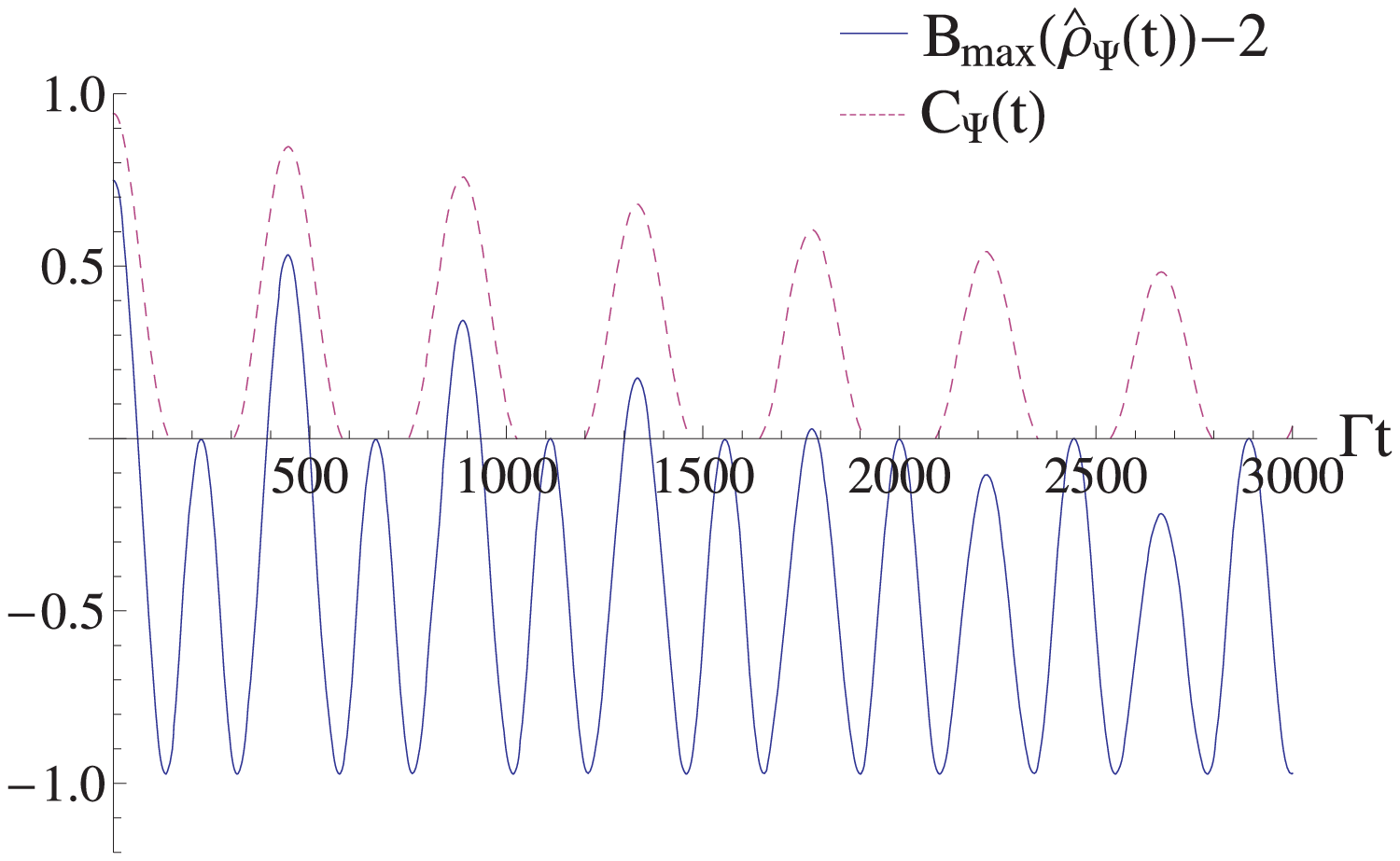}}
\end{center}
\caption{\label{fig:BmaxandC}Maximum of Bell function, $B_\mathrm{max}-2$ (solid line) and concurrence $C$ (dashed line) as a function of the dimensionless time $\Gamma t$ for initial Bell-like states $\ket{\Phi}=(\ket{01}+\sqrt{2}\ket{10})\sqrt{3}$ (upper plot)  and $\ket{\Psi}=(\ket{00}+\sqrt{2}\ket{11})/\sqrt{3}$ (lower plot). The value of $\alpha^2$ is fixed at $1/3$ so that dark intervals of entanglement ($C=0$) occur for the state $\ket{\Psi}$. The coupling parameter is fixed so that $\lambda/\Gamma=10^{-4}$ representing strong non-Markovian effects.}
\end{figure}
To quantify entanglement, we adopt the concurrence $C$ \cite{wootters1998PRL}, being $C=0$ for non-entangled states and $C=1$ for maximally entangled states. The time behavior of $B_\mathrm{max}$ and of $C$ is plotted, for the same values of the parameters, in Fig.~\ref{fig:BmaxandC}. The plot evidences the appearance of regions of entanglement ($C>0$) where however the CHSH-Bell inequality is not violated ($B_\mathrm{max}\leq2$). In these regions it is therefore not possible to say with certainty that quantum correlations are not reproducible by a classical local model \cite{werner1989PRA}. From the figures one sees that $B_\mathrm{max}$ reaches for the first time the classical threshold value $B_\mathrm{max}=2$ rather earlier than the first entanglement disappearance ($C=0$). This behavior confirms what already found previously in comparing the dynamics of Bell inequality violation with entanglement decay for two qubits subjected to local decoherence in the Markovian limit at zero \cite{miran2004PLA} and finite temperature \cite{kofman2008PRA}. Successively, $B_\mathrm{max}$ remains below 2 until $C$ reaches again a threshold value. In particular, for the initial state $\ket{\Phi}$, we find that this threshold value, for any time, is given by $2\alpha\beta/(1+\alpha^2\beta^2)$; the maximum value of this concurrence threshold is $0.8$ and it occurs when the state is initially maximally entangled ($\beta=1/\sqrt{2}$), while in the case represented in Fig.~\ref{fig:BmaxandC} its value is $\approx0.77$. For the initial state $\ket{\Psi}$, the analytical expression of the corresponding threshold value of concurrence is more complex and we do not report it; its maximum value is $\approx0.60$ and it is obtained for $\beta\approx0.79$, while for the case considered in Fig.~\ref{fig:BmaxandC} it takes the value $\approx0.59$. We have then times of high values of concurrence when the CHSH-Bell inequality is not violated. We also point out that the peaks of $B_\mathrm{max}$ decay much more quickly than the ones of $C$.

\begin{figure}
\begin{center}
{\includegraphics[width=0.3\textwidth, height=0.13\textheight]{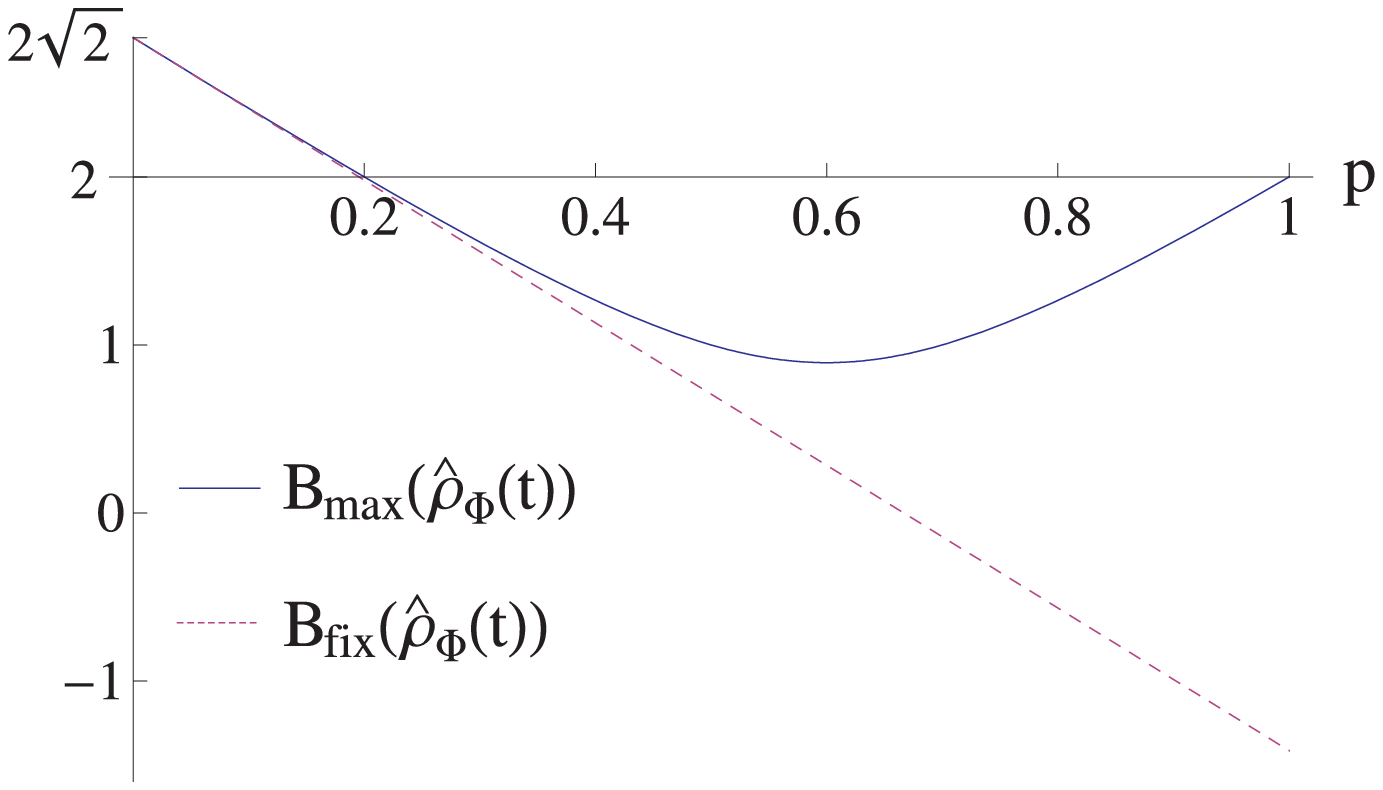}\vspace{0.5cm}
\includegraphics[width=0.3\textwidth, height=0.13\textheight]{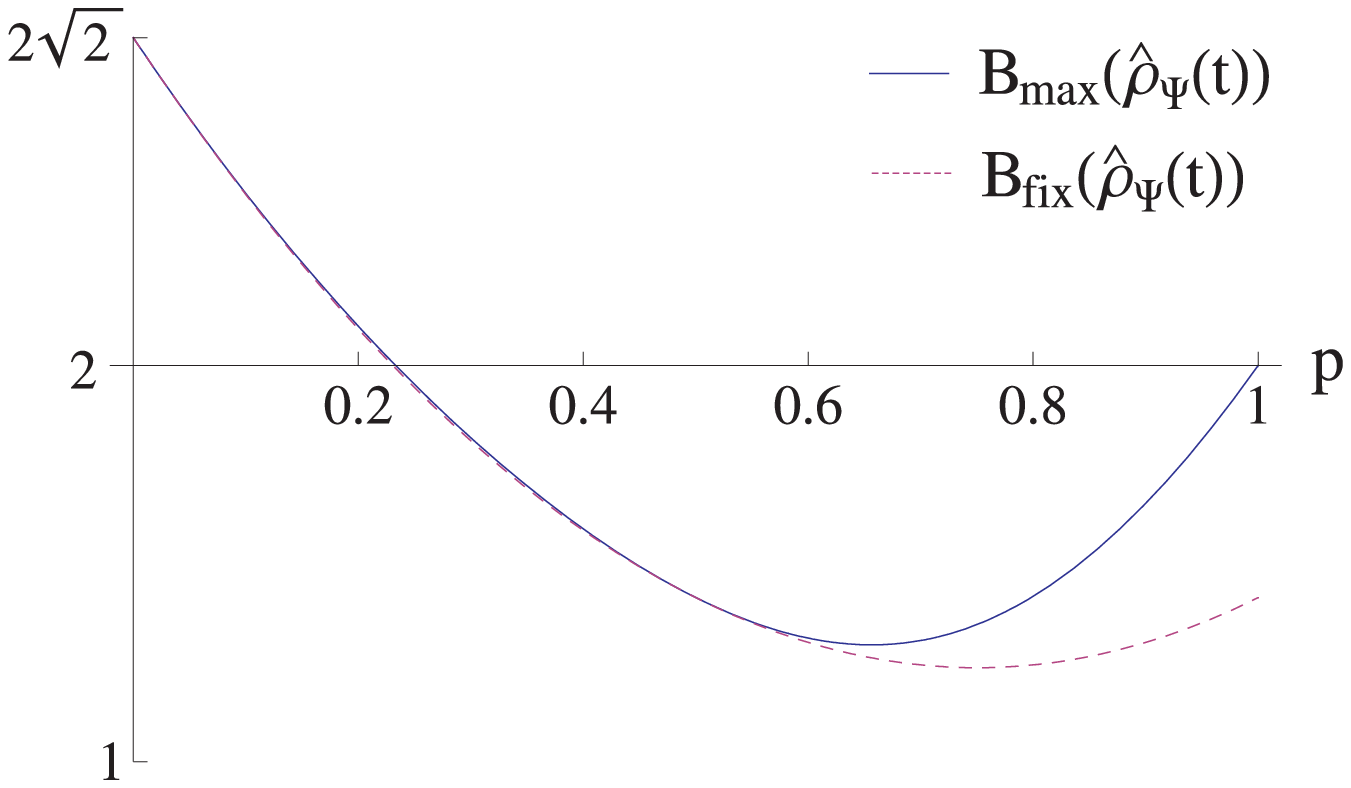}}
\end{center}
\caption{\label{fig:BmaxBfix}Comparison between the maximum of Bell function $B_\mathrm{max}$ and the fixed angles function $B_\mathrm{fix}$ in terms of the quantum channel decay probability $p$, for initial Bell states $\ket{\Phi}=(\ket{01}\pm\ket{10})/\sqrt{2}$ (upper plot) and $\ket{\Psi}=(\ket{00}\pm\ket{11})/\sqrt{2}$ (lower plot).}
\end{figure}

In the analysis above, the revivals of INLCs have been analyzed in terms of $B_\textrm{max}$, which is obtained by using the time dependent angles $\theta_2(t)$ and $\theta'_2(t)$. We however will show that, by suitably fixing these angles, the relative difference between the Bell function at fixed angles, $B_\mathrm{fix}$, and the maximum $B_\mathrm{max}$, is in practice negligible when both are above the classical threshold $B=2$. To this aim, we fix the angles so that $B_\mathrm{fix}$ coincides with $B_\mathrm{max}$ at $t=0$. For both the initial states $\hat{\rho}_\Phi(0)=\ket{\Phi}\bra{\Phi}$ and $\hat{\rho}_\Psi(0)=\ket{\Psi}\bra{\Psi}$ this happens for $\theta_2(0)=\pi-\theta'_2(0)=\arctan(2\alpha\beta)$, which leads to
\begin{eqnarray}\label{Bellfunctionfixed}
   B_\mathrm{fix}(\hat{\rho}_\Phi(t))&=&\frac{2}{\sqrt{1+4\alpha^2\beta^2}}[1+4\alpha^2\beta^2-(2+4\alpha^2\beta^2)p],\nonumber\\
   B_\mathrm{fix}(\hat{\rho}_\Psi(t))&=&\frac{2}{\sqrt{1+4\alpha^2\beta^2}}[1+4\alpha^2\beta^2-4\beta^2(\alpha^2+1)p\nonumber\\
   &&\hspace{2.1cm}+4\beta^2p^2].
\end{eqnarray}
These two quantities, for maximally entangled initial states, are plotted in Fig.~\ref{fig:BmaxBfix} with the corresponding $B_\mathrm{max}(\hat{\rho}_\Phi(t))$ and $B_\mathrm{max}(\hat{\rho}_\Psi(t))$ as functions of the decay probability $p$.
By inspection, for both states, $B_\mathrm{fix}\approx B_\mathrm{max}$ above the classical threshold $B=2$, the relative error $\delta_\mathrm{r}=(B_\mathrm{max}-B_\mathrm{fix})/B_\mathrm{max}$ falling in the range $0\leq\delta_\mathrm{r}\leq10^{-3}$. A realization of the dynamical Bell test that experimentally reveals ``Bell islands'' is thus possible, for any practical purpose, just by following $B_\mathrm{fix}$, with the practical advantage of not changing the angles with time.

Conditions for observing the presence of entanglement revivals are within the range of current experimental capabilities as encountered in solid state devices or in cavity quantum electrodynamics \cite{bellomo2007PRL,devega2005PRA,haroche2006book}. Conditions required for observing revivals of INLCs are more extreme and presumably, for the moment, non obtainable within these contexts. However, a recent optical experiment, exploiting a Sagnac-like interferometer that simulates the amplitude decay channel with the decay probability $p$ linked to the rotation angle of an half-wave plate \cite{almeida2007Science}, has made possible to follow the two-qubit entanglement dynamics for a form of $p$ corresponding to Markovian reservoirs. As a matter of fact, this experimental procedure is valid for any $p$, independently from the values of parameters, so that by taking the explicit form of Eq.~(\ref{decayprobability}) one can in principle obtain, with the same setup, the time evolution of concurrence also for non-Markovian environments, that is also for the dashed curves plotted in Fig.\ref{fig:BmaxandC}. In this optical context, the qubit states $\ket{0}$, $\ket{1}$ are coded by the $H$ (horizontal) and $V$ (vertical) polarizations of a photon, while the environment states are represented by two different momentum modes of the photon and pure entangled photon pairs are repeatedly generated for each value of the decay probability $p$ \cite{almeida2007Science}.

Here we suggest that, with appropriate modifications, the same experimental setup may also be used to follow the revivals of INLCs, overcoming the problem of reaching, in the previous contexts, the required physical conditions. This realization of the dynamical Bell test could be accomplished by substituting, in the optical setup previously described, the final detectors with a standard Bell analyzer \cite{kwiat2001Nature}. The orientation angle of each polarizer of the Bell analyzer, $\theta_\textrm{p}$, corresponds to the half angle $\theta$ of the unit vector $\textbf{O}$, $\theta_\mathrm{p}=\theta/2$ \cite{nielsenchuang}. The photon polarization measurements have to be performed with the appropriate settings of the polarizer angles. In particular, the initial states of the qubits with $\alpha=\beta=1/\sqrt{2}$ are coded into $\ket{\Phi_+}=(\ket{HV}+\ket{VH})/\sqrt{2}$ or $\ket{\Psi_+}=(\ket{HH}+\ket{VV})/\sqrt{2}$ and the polarizers must correspondingly be set at the standard angles $\{\theta_{\mathrm{p}1},\theta'_{\mathrm{p}1},\theta_{\mathrm{p}2},\theta'_{\mathrm{p}2}\}=\{0,45^\circ,22.5^\circ,67.5^\circ\}$\cite{clauser}; while the initial states with $\alpha=-\beta=1/\sqrt{2}$ are $\ket{\Phi_-}=(\ket{HV}-\ket{VH})/\sqrt{2}$ or $\ket{\Psi_-}=(\ket{HH}-\ket{VV})/\sqrt{2}$ and the standard angles are $\{\theta_{\mathrm{p}1},\theta'_{\mathrm{p}1},\theta_{\mathrm{p}2},\theta'_{\mathrm{p}2}\}=\{0,135^\circ,22.5^\circ,67.5^\circ\}$. The realization of this Bell test would check the $p$-dependence of $B_\mathrm{fix}$ given by Eq.~(\ref{Bellfunctionfixed}) and plotted in Fig.~\ref{fig:BmaxBfix}, yielding the observation of the ``Bell Islands'' with $p$ given by Eq.~(\ref{decayprobability}).

In conclusion, we have studied the dynamics of CHSH-Bell inequality violations for two independent qubits, each embedded in an environment with memory, and compared it with the dynamics of entanglement, as measured by concurrence. We have individuated the time regions when the quantum correlations of the two qubits are not reproducible with certainty by a classical local hidden variable model. We have shown that there exist time regions when the CHSH-Bell inequality is not violated even in correspondence to high values of entanglement (up to $C\approx0.8$). We have also suggested that both dynamical behaviors here determined could be observed by adopting a modification of a recent optical experimental setup. The results here found indicate that the entanglement, even for rather high values of concurrence, cannot be exploited with certainty in those circumstances where the non classical reproducibility of quantum correlations is essential, as, e.g., in quantum cryptography.

R.L.F. (G.C.) acknowledges partial support by MIUR project II04C0E3F3 (II04C1AF4E) \textit{Collaborazioni Interuniversitarie ed Internazionali tipologia C}.


\begin{thebibliography}{24}
\expandafter\ifx\csname natexlab\endcsname\relax\def\natexlab#1{#1}\fi
\expandafter\ifx\csname bibnamefont\endcsname\relax
  \def\bibnamefont#1{#1}\fi
\expandafter\ifx\csname bibfnamefont\endcsname\relax
  \def\bibfnamefont#1{#1}\fi
\expandafter\ifx\csname citenamefont\endcsname\relax
  \def\citenamefont#1{#1}\fi
\expandafter\ifx\csname url\endcsname\relax
  \def\url#1{\texttt{#1}}\fi
\expandafter\ifx\csname urlprefix\endcsname\relax\def\urlprefix{URL }\fi
\providecommand{\bibinfo}[2]{#2}
\providecommand{\eprint}[2][]{\url{#2}}

\bibitem[{\citenamefont{Nielsen and Chuang}(2000)}]{nielsenchuang}
\bibinfo{author}{\bibfnamefont{M.~A.} \bibnamefont{Nielsen}} \bibnamefont{and}
  \bibinfo{author}{\bibfnamefont{I.~L.} \bibnamefont{Chuang}},
  \emph{\bibinfo{title}{Quantum Computation and Quantum Information}}
  (\bibinfo{publisher}{Cambridge University Press}, \bibinfo{year}{2000}).

\bibitem[{\citenamefont{Bennett and DiVincenzo}(2000)}]{bennett2000Nature}
\bibinfo{author}{\bibfnamefont{C.~H.} \bibnamefont{Bennett}} \bibnamefont{and}
  \bibinfo{author}{\bibfnamefont{D.~P.} \bibnamefont{DiVincenzo}},
  \bibinfo{journal}{Nature} \textbf{\bibinfo{volume}{404}},
  \bibinfo{pages}{247} (\bibinfo{year}{2000}).

\bibitem[{\citenamefont{Yu and Eberly}(2004)}]{yu2004PRL}
\bibinfo{author}{\bibfnamefont{T.}~\bibnamefont{Yu}} \bibnamefont{and}
  \bibinfo{author}{\bibfnamefont{J.~H.} \bibnamefont{Eberly}},
  \bibinfo{journal}{Phys. Rev. Lett.} \textbf{\bibinfo{volume}{93}},
  \bibinfo{pages}{140404} (\bibinfo{year}{2004}).

\bibitem[{\citenamefont{Bellomo et~al.}(2007)\citenamefont{Bellomo, {Lo
  Franco}, and Compagno}}]{bellomo2007PRL}
\bibinfo{author}{\bibfnamefont{B.}~\bibnamefont{Bellomo}},
  \bibinfo{author}{\bibfnamefont{R.}~\bibnamefont{{Lo Franco}}},
  \bibnamefont{and} \bibinfo{author}{\bibfnamefont{G.}~\bibnamefont{Compagno}},
  \bibinfo{journal}{Phys. Rev. Lett.} \textbf{\bibinfo{volume}{99}},
  \bibinfo{pages}{160502} (\bibinfo{year}{2007}).

\bibitem[{\citenamefont{Bellomo et~al.}(2008)\citenamefont{Bellomo, {Lo
  Franco}, and Compagno}}]{bellomo2008PRA}
\bibinfo{author}{\bibfnamefont{B.}~\bibnamefont{Bellomo}},
  \bibinfo{author}{\bibfnamefont{R.}~\bibnamefont{{Lo Franco}}},
  \bibnamefont{and} \bibinfo{author}{\bibfnamefont{G.}~\bibnamefont{Compagno}},
  \bibinfo{journal}{Phys. Rev. A} \textbf{\bibinfo{volume}{77}},
  \bibinfo{pages}{032342} (\bibinfo{year}{2008}).

\bibitem[{\citenamefont{Maniscalco et~al.}(2008)\citenamefont{Maniscalco,
  Francica, Zaffino, {Lo Gullo}, and Plastina}}]{maniscalco2008PRL}
\bibinfo{author}{\bibfnamefont{S.}~\bibnamefont{Maniscalco}},
  \bibinfo{author}{\bibfnamefont{F.}~\bibnamefont{Francica}},
  \bibinfo{author}{\bibfnamefont{R.~L.} \bibnamefont{Zaffino}},
  \bibinfo{author}{\bibfnamefont{N.}~\bibnamefont{{Lo Gullo}}},
  \bibnamefont{and} \bibinfo{author}{\bibfnamefont{F.}~\bibnamefont{Plastina}},
  \bibinfo{journal}{Phys. Rev. Lett.} \textbf{\bibinfo{volume}{100}},
  \bibinfo{pages}{090503} (\bibinfo{year}{2008}).

\bibitem[{\citenamefont{Bellomo et~al.}()\citenamefont{Bellomo, {Lo Franco},
  and Compagno}}]{bellomo2008arxiv}
\bibinfo{author}{\bibfnamefont{B.}~\bibnamefont{Bellomo}},
  \bibinfo{author}{\bibfnamefont{R.}~\bibnamefont{{Lo Franco}}},
  \bibnamefont{and} \bibinfo{author}{\bibfnamefont{G.}~\bibnamefont{Compagno}},
  \bibinfo{note}{preprint quant-ph/0805.3056}.

\bibitem[{\citenamefont{Wang et~al.}()\citenamefont{Wang, Zhang, and
  Liang}}]{wang2008arxiv}
\bibinfo{author}{\bibfnamefont{F.-Q.} \bibnamefont{Wang}},
  \bibinfo{author}{\bibfnamefont{Z.-M.} \bibnamefont{Zhang}}, \bibnamefont{and}
  \bibinfo{author}{\bibfnamefont{R.-S.} \bibnamefont{Liang}},
  \bibinfo{note}{preprint quant-ph/0805.2876}.

\bibitem[{\citenamefont{Werner}(1989)}]{werner1989PRA}
\bibinfo{author}{\bibfnamefont{R.~F.} \bibnamefont{Werner}},
  \bibinfo{journal}{Phys. Rev. A} \textbf{\bibinfo{volume}{40}},
  \bibinfo{pages}{4277} (\bibinfo{year}{1989}).

\bibitem[{\citenamefont{Masanes et~al.}(2008)\citenamefont{Masanes, Liang, and
  Doherty}}]{masanes2008PRL}
\bibinfo{author}{\bibfnamefont{L.}~\bibnamefont{Masanes}},
  \bibinfo{author}{\bibfnamefont{Y.-C.} \bibnamefont{Liang}}, \bibnamefont{and}
  \bibinfo{author}{\bibfnamefont{A.~C.} \bibnamefont{Doherty}},
  \bibinfo{journal}{Phys. Rev. Lett.} \textbf{\bibinfo{volume}{100}},
  \bibinfo{pages}{090403} (\bibinfo{year}{2008}).

\bibitem[{\citenamefont{Acin et~al.}(2006)\citenamefont{Acin, Gisin, and
  Masanes}}]{acin2006PRL}
\bibinfo{author}{\bibfnamefont{A.}~\bibnamefont{Acin}},
  \bibinfo{author}{\bibfnamefont{N.}~\bibnamefont{Gisin}}, \bibnamefont{and}
  \bibinfo{author}{\bibfnamefont{L.}~\bibnamefont{Masanes}},
  \bibinfo{journal}{Phys. Rev. Lett.} \textbf{\bibinfo{volume}{97}},
  \bibinfo{pages}{120405} (\bibinfo{year}{2006}).

\bibitem[{\citenamefont{Gisin and Thew}(2007)}]{gisin2007natphoton}
\bibinfo{author}{\bibfnamefont{N.}~\bibnamefont{Gisin}} \bibnamefont{and}
  \bibinfo{author}{\bibfnamefont{R.}~\bibnamefont{Thew}},
  \bibinfo{journal}{Nature Photon.} \textbf{\bibinfo{volume}{1}},
  \bibinfo{pages}{165} (\bibinfo{year}{2007}).

\bibitem[{\citenamefont{Bell}(1964)}]{bell}
\bibinfo{author}{\bibfnamefont{J.~S.} \bibnamefont{Bell}},
  \bibinfo{journal}{Physics} \textbf{\bibinfo{volume}{1}}, \bibinfo{pages}{195}
  (\bibinfo{year}{1964}).

\bibitem[{\citenamefont{Clauser et~al.}(1969)\citenamefont{Clauser, Horne,
  Shimony, and Holt}}]{clauser}
\bibinfo{author}{\bibfnamefont{J.~F.} \bibnamefont{Clauser}},
  \bibinfo{author}{\bibfnamefont{M.~A.} \bibnamefont{Horne}},
  \bibinfo{author}{\bibfnamefont{A.}~\bibnamefont{Shimony}}, \bibnamefont{and}
  \bibinfo{author}{\bibfnamefont{R.~A.} \bibnamefont{Holt}},
  \bibinfo{journal}{Phys. Rev. Lett.} \textbf{\bibinfo{volume}{23}},
  \bibinfo{pages}{880} (\bibinfo{year}{1969}).

\bibitem[{\citenamefont{Gisin}(1991)}]{gisin1991PLA}
\bibinfo{author}{\bibfnamefont{N.}~\bibnamefont{Gisin}},
  \bibinfo{journal}{Phys. Lett. A} \textbf{\bibinfo{volume}{154}},
  \bibinfo{pages}{201} (\bibinfo{year}{1991}).

\bibitem[{\citenamefont{Breuer and Petruccione}(2002)}]{petru}
\bibinfo{author}{\bibfnamefont{H.-P.} \bibnamefont{Breuer}} \bibnamefont{and}
  \bibinfo{author}{\bibfnamefont{F.}~\bibnamefont{Petruccione}},
  \emph{\bibinfo{title}{The Theory of Open Quantum Systems}}
  (\bibinfo{publisher}{Oxford University Press}, \bibinfo{address}{Oxford, New
  York}, \bibinfo{year}{2002}).

\bibitem[{\citenamefont{Horodecki et~al.}(1995)\citenamefont{Horodecki,
  Horodecki, and Horodecki}}]{horodecki1995PLA}
\bibinfo{author}{\bibfnamefont{M.}~\bibnamefont{Horodecki}},
  \bibinfo{author}{\bibfnamefont{P.}~\bibnamefont{Horodecki}},
  \bibnamefont{and}
  \bibinfo{author}{\bibfnamefont{R.}~\bibnamefont{Horodecki}},
  \bibinfo{journal}{Phys. Lett. A} \textbf{\bibinfo{volume}{200}},
  \bibinfo{pages}{340} (\bibinfo{year}{1995}).

\bibitem[{\citenamefont{Wootters}(1998)}]{wootters1998PRL}
\bibinfo{author}{\bibfnamefont{W.~K.} \bibnamefont{Wootters}},
  \bibinfo{journal}{Phys. Rev. Lett.} \textbf{\bibinfo{volume}{80}},
  \bibinfo{pages}{2245} (\bibinfo{year}{1998}).

\bibitem[{\citenamefont{Miranowicz}(2004)}]{miran2004PLA}
\bibinfo{author}{\bibfnamefont{A.}~\bibnamefont{Miranowicz}},
  \bibinfo{journal}{Phys. Lett. A} \textbf{\bibinfo{volume}{327}},
  \bibinfo{pages}{272} (\bibinfo{year}{2004}).

\bibitem[{\citenamefont{Kofman and Korotkov}(2008)}]{kofman2008PRA}
\bibinfo{author}{\bibfnamefont{A.~G.} \bibnamefont{Kofman}} \bibnamefont{and}
  \bibinfo{author}{\bibfnamefont{A.~N.} \bibnamefont{Korotkov}},
  \bibinfo{journal}{Phys. Rev. A} \textbf{\bibinfo{volume}{77}},
  \bibinfo{pages}{052329} (\bibinfo{year}{2008}).

\bibitem[{\citenamefont{de~Vega et~al.}(2005)\citenamefont{de~Vega, Alonso, and
  Gaspard}}]{devega2005PRA}
\bibinfo{author}{\bibfnamefont{I.}~\bibnamefont{de~Vega}},
  \bibinfo{author}{\bibfnamefont{D.}~\bibnamefont{Alonso}}, \bibnamefont{and}
  \bibinfo{author}{\bibfnamefont{P.}~\bibnamefont{Gaspard}},
  \bibinfo{journal}{Phys. Rev. A} \textbf{\bibinfo{volume}{71}},
  \bibinfo{pages}{023812} (\bibinfo{year}{2005}).

\bibitem[{\citenamefont{Haroche and Raimond}(2006)}]{haroche2006book}
\bibinfo{author}{\bibfnamefont{S.}~\bibnamefont{Haroche}} \bibnamefont{and}
  \bibinfo{author}{\bibfnamefont{J.~M.} \bibnamefont{Raimond}},
  \emph{\bibinfo{title}{Exploring the Quantum: Atoms, Cavities, and Photons}}
  (\bibinfo{publisher}{Oxford University Press, USA}, \bibinfo{address}{Oxford,
  New York}, \bibinfo{year}{2006}).

\bibitem[{\citenamefont{Almeida et~al.}(2007)}]{almeida2007Science}
\bibinfo{author}{\bibfnamefont{M.~P.} \bibnamefont{Almeida}}
  \bibnamefont{et~al.}, \bibinfo{journal}{Science}
  \textbf{\bibinfo{volume}{316}}, \bibinfo{pages}{579} (\bibinfo{year}{2007}).

\bibitem[{\citenamefont{Kwiat et~al.}(2001)\citenamefont{Kwiat, Barraza-Lopez,
  Stefanov, and Gisin}}]{kwiat2001Nature}
\bibinfo{author}{\bibfnamefont{P.~G.} \bibnamefont{Kwiat}},
  \bibinfo{author}{\bibfnamefont{S.}~\bibnamefont{Barraza-Lopez}},
  \bibinfo{author}{\bibfnamefont{A.}~\bibnamefont{Stefanov}}, \bibnamefont{and}
  \bibinfo{author}{\bibfnamefont{N.}~\bibnamefont{Gisin}},
  \bibinfo{journal}{Nature} \textbf{\bibinfo{volume}{409}},
  \bibinfo{pages}{1014} (\bibinfo{year}{2001}).

\end{thebibliography}
\end{document}